\documentclass[aps,pra,reprint,showpacs,showkeys,groupedaddress,twocolumn, 10pt]{revtex4-1}
\usepackage{amsmath, amssymb}
\usepackage{graphicx}
\usepackage{subfigure}
\usepackage{color}
\newcommand*{\rom}[1]{}

\begin{document}

\title{Effect of non-local interactions on the vortex solution in Bose-Einstein Condensates}

\author{Abhijit Pendse} 
\thanks{abhijeet.pendse@students.iiserpune.ac.in} 
\author {A. Bhattacharyay} 
\thanks{a.bhattacharyay@iiserpune.ac.in}
\address{Department of Physics, Indian Institute of Science Education and Research, Pune, Maharashtra 411008, India.}


\date{\today}

\begin{abstract}
We consider the Gross-Pitaevskii (GP) model of a Bose-Einstein Condensate (BEC) to study a single vortex line in the presence of non-local repulsive s-wave scattering. We show that in addition to the vortex solution with core width of the order of the healing length, there exists a vortex solution whose width is a microscopic length scale of the order of s-wave scattering length and is independent of the healing length. We compare the two classes of vortex solution and show the region where one can possibly observe the vortex whose width is of the order of scattering length.
\end{abstract}




\maketitle


\section{Introduction} 
Healing length of a Bose-Einstein Condensate (BEC) is the distance from a hard-wall boundary at which a BEC heals to its ground state uniform density. There exists selection of this length scale (healing length $\xi_{0}$) via an exact solution of the full non-linear Gross-Pitaevskii (GP) equation in one dimension (1D)\cite{ps,pethick}. Note that, a vortex neither involves a hard boundary nor is it a structure in 1D. Due to these reasons one might expect to get another length scale for the size of a vortex core apart from the healing length when a non-local correction to the local GP equation is present.

\par
The conventional vortex solution of core size $\xi_{0}$ is in general arrived at within the realm of a local GP equation where one takes into account contact interactions between bosons. The three body interactions are suppressed in an atomic BEC by working in the diluteness limit $a<<n^{-\frac{1}{3}}$. Even in a dilute condensate, however, if one has to probe a vortex line with a core so thin that the Lindemann criterion works\cite{vortex_lindemann_cooper, vortex_lindemann_rozhkov}, one actually probes a length scale of the order of scattering length $a$. In this situation the interaction cannot be treated as a contact interaction and one should take corrections for the non-locality of interactions.

\par
In this paper we take into account the leading order non-local interaction correction on top of the local GP dynamics to study here a single vortex in an open condensate. In this procedure of considering corrections to the local GP dynamics we use first a standard correction term as employed to the local GP equation considering the microscopics of the non-local nature of inter-atomic interactions in the references\cite{nonlocal_wang, nonlocal_pethick} and then propose a generalization using a Taylor expansion of the interaction term. We compare the energy of this new class of vortices with the conventional ones and propose a possible way to experimentally realize this new class of vortices which has core-size of the order of scattering length.

\par
The non-locality of interactions induced vortices of core-size of the order of scattering length (thin vortex) which we show in the present paper are obviously of higher energy than the vortices of core-size of the order of healing length (thick vortex). Therefore, the class of the thin vortices, we capture here in the presence of non-local corrections, can only show up in a situation where the thick vortex solution breaks down. It is well known that the thick vortex solution which is arrived at variationally on the basis of an approximate form of solution (so called Pad\'e approximation) exists for $\xi_0<< D$ where $D$ is a large distance cut off employed in the variational calculations in order to manage the logarithmic divergence of energy of the vortex. In this paper we also give a plausibility argument as to how to reach the $\xi_{0}\sim D$ limit in a realistic situation where the conventional thick vortex solution would break down and one might experimentally capture the thin vortex solution that we obtain. Such a scenario, if exists, can possibly result in vortex lattice melting which is of immense importance in the context of atomic quantum Hall effect which is yet to be experimentally realized\cite{vortex_quantum_review_susanne, vortex_quantum_regnault, vortex_ultrafast_dalibard, vortex_rashba_nikoli, vortex_quantum_rougerie, vortex_quantum_furukawa}. 

\par
The paper is organized in the following way. We first discuss the well-known thick vortex solution using the standard procedure in the dilute limit of the GP equation and show that for the existence of this solution, or for the demand of the variational method that captures this solution, the size of the core of this solution is $\xi_0 << D$. The variational method will break down for $\xi_0 \simeq D$. Then we show that taking the leading order correction for non-local interactions on top of the GP dynamics at the dilute limit does even not help this same class of solution survive in the regime $\xi_0 \simeq D$. This is something important to check because the non-local correction actually introduces a competing length scale in the dynamics which is of the order of the s-wave scattering length $a$ and one would expect to get a solution here reflecting this length scale. We capture this solution (thin vortex) using similar variational method in the following section using a model already in literature which provides standard leading order non-local correction to GP equation. We show that, unlike the thick vortices, the existence of this new class of solution does not depend on the radial cut off $D$ and, therefore, can exist in the regime $\xi_0\simeq D$ where the thick vortex solution breaks down. We then generalize results to vortices of any quantum of circulation which is followed by a detailed discussion of our results in the context of possibility of vortex lattice melting using such thin vortices.

\section{Standard Theory} 
Let us look at how \cite{ps} one gets a standard vortex solution in local GP model. It would be explicitly shown that the length scale selection is applicable only when $D>>\xi_{0}$, where $D$ is the radial cut off which is roughly the lateral extent allowed to a single vortex. A radial cut off for a vortex energy calculation is essential because the energy of the vortex increases logarithmically with its radial spread. This crucial condition shall be utilized, in what follows, for showing a comparison between the two classes of vortex solutions. The theory that is shown in this section is quite standard and well known, except for the part presented in subsection C.


\subsection{Solution from GP dynamics}

 The local Gross-Pitaevskii (GP) equation at dilute limit $a<<n^{-1/3}$, where $n$ is density of the condensate, is given by

\begin{equation}
\begin{split}
i\hbar\frac{\partial\psi_{0}({\bf{r}},t)}{\partial t}= \Big(-\frac{\hbar ^{2}}{2m} \nabla^{2}&+V_{ext}({\bf{r}},t)\\
&+ g\;|\psi_{0}({\bf{r}},t)|^{2} \Big)\psi_{0}({\bf{r}},t) .
\end{split}
\label{eq:contact}
\end{equation}

Here $\psi_{0}({\bf{r}},t)$ is the BEC order parameter.The particles of BEC interact here via contact interactions ($\delta-$function interaction potential). $V_{ext}({\textbf{r}},t)$ is the external trapping potential applied to the BEC. $|\psi_{0}({\textbf{r}},t)|^{2}$ is the density $n$ of the BEC. Taking the ansatz $\psi_{0}({\textbf{r}},t)=\psi_{0}({\textbf{r}}) e^{-i\mu t/\hbar}$ in presence of a time-independent external potential, where $\mu$ is the chemical potential, one gets a time independent form of Eq.(\ref{eq:contact}) as

\begin{equation}
 \Big(-\frac{\hbar ^{2}}{2m} \nabla^{2}+V_{ext}({\bf{r}})-\mu+ g\;|\psi_{0}({\bf{r}})|^{2} \Big)\psi_{0}({\bf{r}})=0 .
\label{eq:time_independent}
\end{equation}

A vortex solution corresponds to $\psi_{0}({\textbf{r}})=e^{is\phi}|\psi_{0}({\textbf{r}})|$. This particular wave-function (order parameter) represents a rotation around the axis of symmetry (say z-axis) with a tangential velocity at a distance $r$ from the vortex-core $v_{s}=\hbar s/mr$, where over a closed contour around z-axis $\oint{\vec{v_{s}}. \vec{dl}}=2\pi\hbar s/m$. Here $s$ is the number of quantum of angular momentum ($h$) the vortex carries. For practical purposes, $s=1$ state is important because it is the most stable one and higher order vortices quickly break into the $s=1$ state.

\par
The radial part of the vortex state is found out from Eq.(\ref{eq:time_independent}) by taking an ansatz $|\psi _{0}|=\sqrt{n} f(\eta)$, where $\eta=r/\xi_{0}$ is the radial distance in the unit of healing length. Thus the equation for $f(\eta)$ $\equiv f$ is

\begin{equation}
 \frac{1}{\eta} \frac{d}{d\eta} \big(\eta\frac{d}{d\eta} f\big)+ \big(1-\frac{s^{2}}{\eta^{2}}\big)f- f^{3}=0 ,
\label{eq:original_vortex}
\end{equation}

where $V_{ext}=0$ has been set for the sake of simplicity to study a single vortex somewhere near the middle of the trap where effects of boundary are not appreciable. 
\par
Far away from the core of the vortex, i.e. at large enough $\eta$, a solution constant over space should exist and one gets non-trivial solution $f=1$ by equating $(f-f^{3})$ to zero. At a small $\eta$, one takes $f\sim\eta^{|s|}$, where $s$ is an integer. Then, Eq.(\ref{eq:original_vortex}) reduces to

\begin{equation*}
|s|^{2}\eta^{|s|-2}-s^{2}\eta^{|s|-2}+ \eta^{|s|}-\eta^{3|s|}=0.
\end{equation*} 

A balance of the dominant terms at small $\eta$ , i.e. $\eta^{|s|-2}$ shows that  $f=\eta^{|s|}$ is a solution and this solution holds for all $s$ at the leading order. For a vortex with $s=1$, the cancellation happens between the terms $ \frac{1}{\eta} \frac{d}{d\eta} f$ and $\frac{1}{\eta^{2}}f$ which is somewhat a special case giving the same solution $f=\eta^{|s|}$.

\par
Since the vortex core solution of any order comes from the leading order terms resulting from the Laplacian, the selection of the scale $\xi_{0}$ is not accomplished in this computation. To actually determine this scale following the present procedure, one has to look at the balance of the terms at next to the leading order. This, however, is not possible if one does not take into account non-local interactions.


\subsection{Scale selection using energy functional}

A scale selection for this vortex core is obtained by the variation of the grand canonical free energy functional of the local GP equation (Eq.1) in general. By taking an ansatz \cite{pethick} for the vortex solution to be of the form $\psi({\bf{r}},t)=\sqrt{n} \: |f(r)| e^{i|s|\phi} e^{-i\mu t/\hbar}$ where $s$ is an integer. Let us consider the case $|s|=1$ and take $|f(r)|=r/\sqrt{\beta^{2}+r^{2}}$ as an ansatz. Since, this is a procedure employing energy variations, the term containing any constant phase factor of $f(r)$ is of no use. We plug this ansatz in the grand canonical energy functional of the local GP equation and minimize the energy with respect to $\beta$ to get a scale selection. The grand canonical energy functional of the local GP equation is \cite{ps}

\begin{equation}
\begin{split}
E=\int_{-\frac{L}{2}}^{\frac{L}{2}}\int_{0}^{2\pi}\int_{0}^{D}&\Big[\frac{\hbar^{2}}{2m}|\nabla\psi({\bf{r}},t)|^{2}\\
& +\frac{g}{2}(|\psi({\bf{r}},t)|-n)^{2} \Big]r\:dr\:d\phi\:dz.
\end{split}
\end{equation}

After putting the aforementioned ansatz for $\psi({\bf{r}},t)$ in the above energy functional, we extremize it by setting $(\partial E/\partial\beta)=0$. This gives the following equation for $\widetilde{\beta}$($\widetilde{\beta}=\beta/D$),

\begin{equation*}
\widetilde{\beta}^{4}[6\xi_{0}^{2}-D^{2}]+\widetilde{\beta}^{2}(4\xi_{0}^{2}-D^{2})+2\xi_{0}^{2}=0,
\end{equation*}
 where $\xi_{0}=\hbar/\sqrt{2mgn}$.
We can find the roots of the above equation in $\widetilde{\beta}$, which gives

\begin{equation*}
\widetilde{\beta}^{2}=\frac{(D^{2}-4\xi_{0}^{2})\pm\sqrt{D^{4}-32\xi_{0}^{4}}}{2(6\xi_{0}^{2}-D^{2})}.
\end{equation*}

For the above equation to have roots, $D^{4}\geq32\xi_{0}^{4}$. This condition is tailored in a BEC while exciting multiple vortices\cite{vortex_dalibard, vortex_ketterle}. In this $D>>\xi_{0}$ limit, we see that taking the $+$ sign makes the numerator $\sim 2D^{2}$. However, we see that there is another factor $[6\xi_{0}^{2}-D^{2}]$ in the denominator, which becomes negative for $D>>\xi_{0}$, in turn making $\widetilde{\beta}^{2}$ negative. If we take the $-$ sign from the $\pm$, the numerator of the above equation becomes $\sim (-\xi_{0}^{2})$ , giving $\widetilde{\beta}^{2}\sim(\xi_{0}^{2}/D^{2})$ for $D>>\xi_{0}$. This gives us a scale selection of $\beta\sim\xi_{0}$. This is the vortex with core size of the order of healing length.

\par
The important point to note here is that for $D^{4}-32\xi_{0}^{4}<0$, the solution breaks down because $\widetilde{\beta}$ becomes complex. Now, $D$ being a free parameter of a single vortex, it can be increased up to the system size, but in a vortex lattice, $D$ is restricted by the position of the nearest neighbours and in what follows we are going to look at this limit after introducing the other class of solution.

\par
\subsection{Existing non-local model}
The condition for the above mentioned class of vortex solution's breakdown appears in the context of a local GP dynamics is what we have seen so far. An important question to ask at this point is - does the leading order non-local correction to the local dynamics improve the situation?  

Let us revisit the above variational procedure taking into consideration non-local interactions. We do that to see if there exists a vortex solution of core size $\xi_{0}$ when non-local interactions are present, despite having $D^{2}<\sqrt{32}\xi_{0}^{2}$. Contemporary modifications to the local GP equation (Eq.(\ref{eq:contact})) \cite{nonlocal_wang, nonlocal_pethick, nonlocal_ketterle} involve adding a first order correction term, considering some microscopic potential underlying s-wave interactions. Using, for example, the notations used by Collin \textit{et al.}, we get the following non-local GP equation \cite{nonlocal_pethick}

\begin{equation}
\begin{split}
i\hbar\frac{\partial\psi_{0}({\bf{r}},t)}{\partial t}&= \Big[-\frac{\hbar ^{2}}{2m} \nabla^{2}+V_{ext}({\bf{r}})\\
&+ g\;\Big(|\psi_{0}({\bf{r}},t)|^{2}+g_{2}\nabla^{2}|\psi_{0}({\bf{r}},t)|^{2}\Big) \Big]\psi_{0}({\bf{r}},t) ,
\end{split}
\label{eq:pethick_1}
\end{equation}
where $g_{2}=\Big(\frac{a^{2}}{3}-\frac{ar_{e}}{2}\Big)$ and $r_{e}$ is the effective range of two-body interaction. Let us take Eq.(\ref{eq:pethick_1}) in the absence of an external potential and determine the nature of vortex solution.

The energy functional for the non-local GP equation given by Eq.(\ref{eq:pethick_1}) is \cite{we}

\begin{equation}
\begin{split}
E=\int_{-\frac{L}{2}}^{\frac{L}{2}}\int_{0}^{2\pi}\int_{0}^{D}&\Big[\frac{\hbar^{2}}{2m}|\nabla\psi({\bf{r}},t)|^{2}+\frac{g}{2}(|\psi({\bf{r}},t)|-n)^{2}+\\
&\frac{gg_{2}}{2}|\psi({\bf{r}},t)|^{2}\nabla^{2}|\psi({\bf{r}},t)|^{2} \Big]r\:dr\:d\phi\:dz.
\end{split}
\end{equation}

We plug in the same ansatz, $|f(r)|=r/\sqrt{\beta^{2}+r^{2}}$ as in the previous section and check for energy functional minimization. Note that, since we do not want three body effects to dominate, it is desirable to work in the diluteness limit $a^{3}n<<1$ and hence $g_{2}<<\xi_{0}^{2}$ which can safely be considered in orders of magnitude as $g_{2}\sim a^{2}$. The minimization condition yields

\begin{equation}
\begin{split}
\widetilde{\beta}^{6}(6\xi_{0}^{2}-D^{2})&+\widetilde{\beta}^{4}(10\xi_{0}^{2}-2D^{2}+4g_{2})\\
&\widetilde{\beta}^{2}(6\xi_{0}^{2}-D^{2}-4g_{2})+2\xi_{0}^{2}=0.
\end{split}
\label{eq:beta_tilde_selection}
\end{equation}

The above condition is cubic in $\widetilde{\beta}^{2}$. The coefficients show that the length scale dictated by $g_{2}$ would be masked by that by $\xi_{0}$, unless $g_{2}\geq \xi_{0}^{2}$. However, this would take us away from the diluteness limit and into the region where three body effects would become important. Hence while we stay in the diluteness limit, the length scale dictated by $\xi_{0}$ would always dominate. Now, in the previous section, we have seen that for $D^{2}<\sqrt{32}\xi_{0}^{2}$, the solution breaks down. For the same condition over most of the parameter space, one can see that all the coefficients of Eq.(\ref{eq:beta_tilde_selection}) are positive which denies any acceptable solution for $\widetilde{\beta}^{2}$. Therefore, even the presence of non-local interaction does not help getting such a vortex solution for $D^{2}<\sqrt{32}\xi_{0}^{2}$ in a system when the leading order non-local correction has been taken into account.

\section{Scale selection from non-local model}
Let us now consider the GP model with non-local correction and fix the length scale of the vortex from the dynamics. This is only doable in the presence of non-local corrections to the local GP equation. We then minimize the energy of this class of vortices by variation and fix the entire profile.

 Considering $\mu=gn$ and $\psi_{0}(\textbf{r})=\sqrt{n}f(r)e^{is\phi}$, we take the standard ansatz, as already mentioned in section A, $f(r)=R^{|s|}$ near the origin, where $R=\beta r$. Using this ansatz, Eq.(\ref{eq:pethick_1}) near the origin is given by 

\begin{equation}
\begin{split}
&\beta^{2} |s|^{2} R^{|s|-2} - \beta^{2} |s|^{2} R^{|s|-2} - 8\pi a n R^{3|s|}\\
&+ 8\pi an R^{|s|} \Big[1-g_{2}\beta^{2}(2|s|)^{2}R^{2|s|-2}\Big]=0,
\end{split}
\label{eq:pethick_2}
\end{equation}

Note that irrespective of the choice of $\beta$ and $|s|$, the balance between the leading order terms, i.e. the first two terms of the above equation is always there. On top of that, due to the non-local correction term introduced in Eq.(\ref{eq:pethick_1}), there now exists a special case of $|s|=1$ where there exists a balance in the sub-leading terms in brackets in Eq.(\ref{eq:pethick_2}). This now imposes a selection of $\beta$, given by $\beta=1/(2\sqrt{g_{2}})$ only for $|s|=1$. This selection exists when $g g_{2}$ is positive, or having $g$ positive when $g_{2}$ is also positive, i.e. $r_{e}<2a/3$. This in general would be a situation when repulsive s-wave scattering is the dominant interaction mechanism in BEC, which is well known to be the case for all metastable BEC. Since, in this regime, $g_{2}\sim a^{2}$, hence, we get from the above $\beta$ selection that $\beta\sim 1/a$. The next obvious line of inquiry would be generalizing this approach to include all vortices. 



\section{Generalization}
The local GP equation follows from its full non-local form

\begin{equation}
\begin{split}
i\hbar\frac{\partial\psi_{0}({\bf{r}},t)}{\partial t}&= \Big(-\frac{\hbar^{2}}{2m} \nabla^{2}+V_{ext}({\bf{r}},t)\Big) \psi_{0}({\bf{r}},t)\\
&+ \psi_{0}({\bf{r}},t) \int{{\bf{dr^{'}}} \psi^{*}_{0}({\bf{r'}},t) V({\bf{r'-r}}) \psi_{0}({\bf{r'}},t)}.
\end{split}
\label{eq:original}
\end{equation}

Let us consider the interaction to be repulsive s-wave scattering, which is the most dominant interaction in a BEC.
\par
Under the first Born approximation, s-wave scattering can be captured by using an effective repulsive soft-potential \cite{ps, sakurai}. The range of this effective repulsive potential is in general of the order of the scattering length. We use this approach in what follows, because the system we deal with is necessarily having lowest energy repulsive s-wave scattering. We Taylor expand the wave-function to explore the structure of $\psi$ and the corresponding density $n$. The effective potential simply sets the length scale over which the density variation is now being probed. Important to note that, although we are capturing here the interactions underlying s-wave scattering through an effective repulsive potential for the sake of simplicity, the actual potential can have bound states and, thus, can undergo Feshbach resonance.
 
\par
We, therefore, take the interaction pseudopotential as $V({\bf{r^{'}-r}})\simeq V_{eff} = [g/(\sqrt{2\pi}a)^{3}]\times exp[-\frac{|{\bf{r-r^{'}}}|^{2}}{2a^{2}}]$. Following the arguments given above, the range of the $V_{eff}$ would be of the order of s-wave scattering length and hence we set the range of $V_{eff}$ as $a$. Doing a Taylor expansion of the wave-function $\psi({\bf{r^{'},t}})$ about $\textbf{r}$, we get

\begin{equation}
\begin{split}
&-\frac{\hbar^{2}}{2m}\frac{1}{r}\frac{d}{dr} \Big(r\frac{d}{dr}|\psi_{0}|\Big)+ \frac{\hbar^{2}s^{2}}{2mr^{2}}|\psi_{0}|+g|\psi_{0}|^{2} \psi_{0}-\mu|\psi_{0}|\\
&+g|\psi_{0}| \Big[ \frac{a^{2}}{2}\frac{1}{r}\frac{d}{dr} \Big(r\frac{d}{dr}|\psi_{0}|^{2}\Big) + \frac{a^{4}}{8}\Big(\frac{1}{r}\frac{d}{dr}\big(r\frac{d}{dr}\big)\Big)^{2}|\psi_{0}|^{2}\\
&+.....+\frac{a^{2l}}{(2l)!!}\Big(\frac{1}{r}\frac{d}{dr}\big(r\frac{d}{dr}\big)\Big)^{2l}|\psi_{0}|^{2}+.... \Big]=0,
\end{split}
\label{eq:modified_vortex}
\end{equation}

\par
Considering $\mu=gn$, $\psi_{0}(\textbf{r})=\sqrt{n}f(r)$, and $f(r)=R^{|s|}$ near the origin, where $R=\beta r$, for a vortex with $|s|$ quanta of circulation, the series in square brackets in Eq.(\ref{eq:modified_vortex}) terminates at $2|s|$-th order in derivatives as all other higher order derivatives would be zero. Thus, we get

\begin{equation*}
\begin{split}
&\beta^{2} |s|^{2} R^{|s|-2} - \beta^{2} |s|^{2} R^{|s|-2} - 8\pi a n R^{3|s|}+ 8\pi an R^{|s|} \Big[1\\
&-\frac{a^{2}\beta^{2}}{2}(2|s|)^{2}R^{2|s|-2}-\frac{a^{4}\beta^{4}}{8}(2|s|)^{2}(2|s|-2)^{2}R^{2|s|-4}\\
&-....-\frac{a^{2|s|}\beta^{2|s|}}{(2|s|)!!}((2|s|)!!)^{2}\Big]=0.
\end{split}
\end{equation*}

The first 3 terms correspond to local GP equation and at the leading order $R^{|s|-2}$ we get the same result as is already shown. The next higher order terms are the first and the last terms in the square bracket which are in balance when $\beta=1/a[(2|s|)!!]^{\frac{1}{2|s|}}$ giving us a selection on $\beta$. Note that, the first term in the square bracket comes from the linear term $\mu|\psi_{0}|$ in Eq.(\ref{eq:modified_vortex}). Note also that, had we considered any other flat symmetric repulsive potential in the place of the Gaussian one, the Taylor expansion would only change coefficients involving $|s|$. Thus, the scaling $\beta\sim 1/a$ is always there.

\par
The natural truncation of the Taylor expansion of the interaction term, depending upon the order of the quantized vortex, is something of immense importance here. There exists no need to truncate the series based on order of magnitude arguments. The structure of the core of a vortex, i.e. $f(r)=R^{|s|}$ naturally truncates the series and that is where the Taylor expansion makes perfect sense to be in use. For the kind of order parameters which does not provide such a natural truncation of the Taylor series, the Taylor expansion method cannot be used for these small scale structures where $\lambda\sim a$.

\par
Having constructed the core with a definite selection of the length scale, we can construct a full variational solution for a vortex with $|s|$ quantum of circulation. To compute the free energy in what follows, we take the energy integral over $R$ varying from $0$ to $D$, where $r=D/\beta$ is the radial cut off. To obtain the solution, we divide the region $(0,D)$ in two parts. These regions are $(0,\alpha_{|s|})$ and $(\alpha_{|s|},D)$. In the region $(0,\alpha_{|s|})$ we consider the vortex solution as $R^{|s|}$ (as was shown above in the analysis to obtain $\beta$) and in the region $(\alpha_{|s|}, D)$, we take the ansatz $f(R)=(1-\lambda_{|s|} e^{-\delta_{|s|} R})$. The choice of this latter part of the solution is particularly good because it rises to unity at large $R$ and it involves two free parameters which we can adjust to get a continuous solution. By matching of these two functions and their derivatives at $R=\alpha_{|s|}$ we obtain $\lambda_{|s|}=(1-(\alpha_{|s|})^{|s|})\cdot exp[|s|(\alpha_{|s|})^{|s|}/(1-(\alpha_{|s|})^{|s|})]$, $\delta_{|s|}=|s|/\big[(\alpha_{|s|})^{(1-|s|)}-\alpha_{|s|}\big]$ as a function of $\alpha_{|s|}$. Then, we plug in the complete solution (from $0$ to $D$) in the energy functional (Eq.(\ref{eq:modified_vortex})) \cite{we} to minimize the energy and thereby fix the position of matching $\alpha_{|s|}$.

\par
The free energy functional corresponding to the general vortex of number of circulation $|s|$ in the solution we have chosen is \cite{we}

\begin{widetext}
\begin{equation}
\begin{split}
E_{v}&= \int_{0}^{D}\frac{2\pi R dR}{\beta^{2}} \Big[\frac{\hbar^{2}\beta^{2}}{2m}n\big(\frac{df}{dR}\big)^{2}+\frac{\hbar^{2}\beta^{2}}{2m}n \frac{f^{2}}{R^{2}}+ \frac{gn^{2}}{2}(1-f^{2})^{2}+\frac{gn^{2}f^{2}}{2} \Big\{\frac{\beta^{2}a^{2}}{2}\frac{1}{R} \frac{d}{dR}\big(R\frac{d}{dR}f^{2} \big)\\
&\hspace{2cm}+\frac{\beta^{4}a^{4}}{8}\Big(\frac{1}{R} \frac{d}{dR}\big(R\frac{d}{dR}\big)\Big)^{2}f^{2}+...+\frac{\beta^{2|s|}a^{2|s|}}{(2|s|)!!}\Big(\frac{1}{R} \frac{d}{dR}\big(R\frac{d}{dR}\big)\Big)^{2|s|}f^{2} \Big\}\Big].
\end{split}
\label{eq:energy_modified}
\end{equation}
\end{widetext}

We use the above mentioned prescription and evaluate the energy minimum in terms of $\alpha_{|s|}$. As we assume the diluteness limit $a^{3}n<<1$, we can take $an<<\beta^{2}$ for any $|s|$ since $\beta \sim 1/a$. Making use of this inequality and keeping terms which go as $\beta^{2}$ over terms which go as $(an)$ we get an expression for free energy as

\begin{equation*}
\begin{split}
E=\frac{2\pi\hbar^{2}n}{m}&\Big[\frac{|s|^{2}}{2}ln\Big(\frac{D}{\alpha_{|s|}}\Big)\\
&+\Big(\frac{\lambda_{|s|}^{2}}{8}\Big)(1+2\alpha_{|s|}\delta_{|s|})e^{-2\alpha_{|s|}\delta_{|s|}} +\frac{|s|\alpha_{|s|}^{2|s|}}{2}\Big].
\end{split}
\end{equation*}


The minimization of the above energy functional gives $\alpha_{|s|}=[(1-|s|+\sqrt{49|s|^{2}-10|s|+1})/(12|s|-2)]^{\frac{1}{|s|}}$. Using this $\alpha_{|s|}$, we can now evaluate $\lambda_{|s|}$ and $\delta_{|s|}$ for a given value of $|s|$, thus we determine $f(R)$ for the interval $(\alpha_{|s|}, D)$.

\par
\begin{figure}[h]
  \includegraphics[width=8cm]{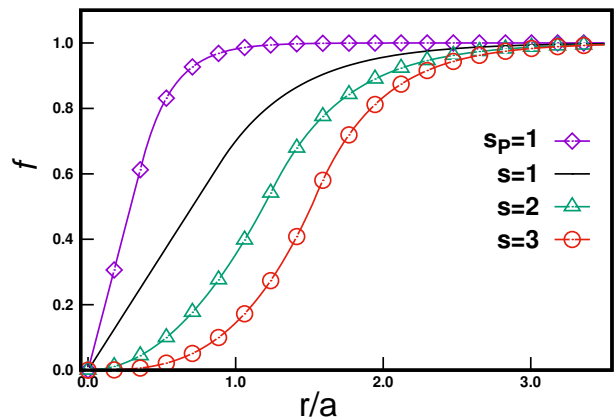}
  \caption{\small {Figure shows the density profile of single vortex line for different circulations $s$ for the generalized model and also for the microscopic interaction model by Collin \textit{et al.} labelled by $s_{P}$ for $|s|=1$ with $r_{e}=a/2$.}}
\label{fig:density_profile}
\end{figure}

We plot $f(R)$ for $|s|=1,2,3$ in Fig.(\ref{fig:density_profile}) for the generalized non-local model and for the non-local model by Collin \textit{et al.} \cite{nonlocal_pethick} with $|s|=1$. For the non-local model of Collin {\it et al} we have taken $r_{e}=a/2$ to plot the density profile of the  vortex. The figure shows that the width of the vortices is of the order of s-wave scattering length.

\section{Thin vortex in a harmonic trap}

Let us now consider the analysis applied above for a BEC in a harmonic trap. We shall discuss a vortex solution with unit circulation, i.e. $s=1$ using the model by Collin \textit{et al.} . We consider cylindrical geometry and assume tight confinement along the $z$ direction. This consideration would give the GP model used by Collin \textit{et al.} as

\begin{equation}
\begin{split}
i\hbar\frac{\partial\psi_{0}({\bf{r}},t)}{\partial t}&= \Big[-\frac{\hbar ^{2}}{2m} \nabla^{2}+\frac{1}{2}m\omega^{2}r^{2}\\
&+ g\;\Big(|\psi_{0}({\textbf{r}},t)|^{2}+g_{2}\nabla^{2}|\psi_{0}({\bf{r}},t)|^{2}\Big) \Big]\psi_{0}({\bf{r}},t) ,
\end{split}
\label{eq:pethick_1_trap}
\end{equation}

where $\omega$ is the trap frequency and $r$ is the radial distance from the trap centre. Due to the harmonic trapping, the ground state of the BEC is no longer one with uniform density, but would fall off away from the centre of the trap. The Thomas-Fermi(TF) density of the condensate can be evaluated by neglecting the kinetic contribution coming from the Laplacian terms. By considering $\psi_{0}=\sqrt{n_{TF}}e^{-i\mu t/\hbar}$, we get the TF density as $n_{TF}=\frac{1}{g}(\mu-\frac{1}{2}m\omega^{2}r^{2})$. We wish to probe the vortex core scaling similar to the previous analysis (where we had considered a uniform background density). To this end, we consider a vortex with unit circulation($s=1$) at a distance $d$ away from the trap centre. If we consider a coordinate system $(\rho,\theta)$ centred at the vortex core, we would get the vortex profile near the core as $\sqrt{n_{TF}(d)}\;\beta\;\rho e^{i\theta}e^{-i\mu t/\hbar}$, where $n_{TF}(d)$ is the TF density at a distance $d$ from the trap centre. In considering a linear profile, we have assumed that the TF density of the condensate remains constant up to the first order in $\rho$. Also, as before we have considered here the vortex as an excited state on top of the ground state density. With respect to the coordinate system $(r,\phi)$ around the trap centre, $\rho$ and $\theta$ transform as $\rho=\sqrt{r^{2}-2rd\cos{\phi}+d^{2}}$ and $\theta=\tan^{-1}[{r\sin{\phi}/(r\cos{\phi}-d)}]$. Considering this, we can write the density profile of the vortex at a distance $d$ from the trap centre as $\psi_{0}({\bf{r}},t)=\sqrt{n_{TF}(d)}\;\beta\;\sqrt{r^{2}-2rd\cos{\phi}+d^{2}}\:e^{i\tan^{-1}[{\frac{r\sin{\phi}}{r\cos{\phi}-d}}]}e^{-i\mu t/\hbar}$. Putting this ansatz in Eq.(\ref{eq:pethick_1_trap}) gives us

\begin{equation}
\begin{split}
\mu&\sqrt{r^{2}-2rd\cos{\phi}+d^{2}}=\frac{\hbar^{2}}{2m}\Big[\frac{e^{i\phi}\sqrt{r^{2}-2rd\cos{\phi}+d^{2}}}{r(d-e^{i\phi}r)}\\
& \hspace{3cm} -\frac{e^{i\phi}\sqrt{r^{2}-2rd\cos{\phi}+d^{2}}}{r(d-e^{i\phi}r)}\Big]\\
& \hspace{2.5cm} +\frac{1}{2}m\omega^{2}r^{2}\sqrt{r^{2}-2rd\cos{\phi}+d^{2}}\\
& \hspace{2cm} +gn_{TF}(d)\beta^{2}(\sqrt{r^{2}-2rd\cos{\phi}+d^{2}})^{3}\\
& \hspace{2cm} +4gg_{2}n_{TF}(d)\beta^{2}\sqrt{r^{2}-2rd\cos{\phi}+d^{2}}.
\end{split}
\label{eq:pethick_2_trap}
\end{equation}

From the equation above, we can see that the terms coming from $\nabla^{2}\psi$ cancel each other out. We then wish to look at the remaining terms in Eq.(\ref{eq:pethick_2_trap}). Since we consider $\rho$ to be small, we go back to the $(\rho,\theta)$ coordinate system. This gives us

\begin{equation*}
\begin{split}
&\rho\Big(\mu-\frac{1}{2}m\omega^{2}d^{2}-gg_{2}n_{TF}(d)\beta^{2}\Big)-\rho^{2}(m\omega^{2}d\cos{\theta})\\
&-\rho^{3}\Big(\frac{1}{2}m\omega^{2}+gn_{TF}(d)\beta^{2}\Big)=0.
\end{split}
\end{equation*}

Up to the leading order in $\rho$ then, we get, [$\mu-\frac{1}{2}m\omega^{2}d^{2}-4gg_{2}n_{TF}(d)\beta^{2}]=0$. As the TF density at a distance $d$ from the trap centre is given as $n_{TF}=\frac{1}{g}(\mu-\frac{1}{2}m\omega^{2}d^{2})$, we get the expression for $\beta$ as $\beta=1/2\sqrt{g_{2}}$, same as for a BEC in the absence of a trap. While, the vortex profile far away from the core may be different from that in the uniform density background case, the scaling of the density profile near the vortex core remains the same. This shows that the thin vortex solution obtained on top of a BEC with uniform density can be obtained for a harmonically trapped BEC as well. An anharmonic term of the order $r^{4}$ in the potential will again get absorbed by the TF density profile and will not change our result up to the scope of the present approximation.

\par

\begin{figure}[h]
  \includegraphics[width=8cm]{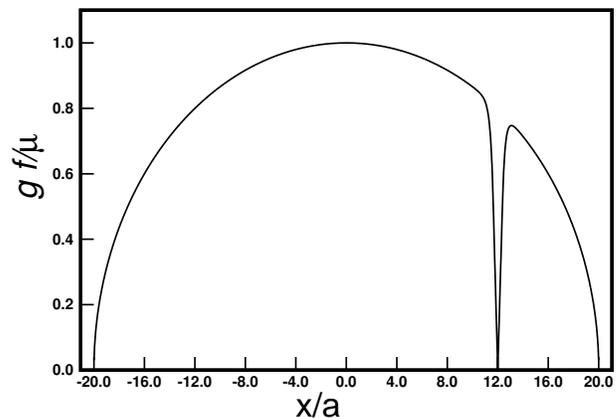}
  \caption{\small {Figure shows the density profile of single vortex line in a harmonically trapped BEC situated at $d/a=12$ from the centre of the trap based on the model by Collin \textit{et al.} for $|s|=1$ with $r_{e}=a/2$.}}
\label{fig:density_profile_harmonic}
\end{figure}

Having obtained the core, we can now use the variational method applied in the previous section to obtain the entire vortex solution. We take the large scale cut-off width for the harmonically trapped BEC to be $D$ and divide the radial interval $(0,D)$ into two regions as before. So long as $D$ is bigger than the width of the actual confinement, there is no problem because the density vanishing beyond the actual width makes the integrand vanish and the integral remains the same as the one obtained considering the actual width. If the vortex is situated at a distance $d$ from the trap centre, there would in effect be three regions in the coordinates of the vortex(along a line joining the trap centre to the core), viz., the region of the core $(-\alpha,\alpha)$ and the regions away from the core $(\alpha,D-d)$ and $(-(D+d),-\alpha)$ . In our calculation, we write the TF density in terms of the radial coordinates of the vortex centre $(\varrho,\theta)$, where $\varrho=\beta\rho$. Thus, the TF density is $n_{TF}(\varrho)=(1/g)(\mu-\frac{m\omega^{2}}{2}[(\varrho^{2}/\beta^{2})+2\varrho (d/\beta)\cos{\theta}+d^{2}])$. We take the ansatz for vortex solution of the form $\psi=f(\varrho)\;e^{i\theta}e^{-i\mu t/\hbar}$. For the interval of the core given by $(0,\alpha)$, $f=\varrho\;n_{TF}(\varrho)$. Outside the core, the ansatz is $f(\varrho)=(1-\lambda e^{-\delta \varrho}) \; n_{TF}(\varrho)$. As before, we match the two solutions and their derivatives at $\varrho=\alpha$. Further, we minimize the energy functional with respect to $\alpha$ as before. Following this procedure, we numerically evaluate the parameters in the variational procedure and thus obtain the profile shown in Fig.(\ref{fig:density_profile_harmonic}) for the constant $m\omega^{2}/2\mu$ set equal to unity. Our variational method works perfectly in determining the large scale profile of the vortex.

\section{Comparison of energy of two classes of vortices}

We now return to the discussion of vortex solution for a uniform BEC in the absence of any trap. Let us now compare the grand canonical energy of the two classes of vortices, one with core size of the order of healing length (thick vortex) and other with core size of the order of s-wave scattering length (thin vortex). To the leading order, the energies of the vortex with core of size of healing length ($E_{\xi_{0}}$) and that with core size of scattering length ($E_{a}$) are given by $E_{\xi_{0}}\sim(\frac{\pi \hbar^{2}Ln}{m}) ln\Big(\frac{|s|D}{\xi_{0}}\Big)$ and $E_{a}\sim(\frac{\pi\hbar^{2}Ln}{m}) ln\Big(\frac{D}{\alpha_{|s|}a}\Big)$ respectively, where $\alpha_{|s|}$ is an order unity number. From these terms, we can see that if the healing length and the s-wave scattering length are of equal order, the two classes of vortices actually become comparable in energy. However, as we change the scattering length and make it smaller by keeping the density fixed, we increase the healing length. In this case, the thick vortex with the length scale corresponding to the healing length will be energetically favoured because it is a state of lower energy.

\par
However, as mentioned earlier that, for the thick vortex solution to exist, the condition is $D^{4}\geq32\xi_{0}^{4}$. In the case $D^{4}<32\xi_{0}^{4}$, the thick vortex would cease to exist as that length scale selection is no longer present. Hence, the vortex solution existing in this regime could be a thin vortex with core size of the order of the s-wave scattering length. 
\par
Let us discuss an experimentally plausible situation where a thick vortex solution does not exist and a thin vortex state can possibly be captured. Note that, the radial cut off $D$ corresponding to a single vortex inside a vortex lattice is rather restricted and is actually much smaller than the system size. In such a situation, one can employ Feshbach mechanism off resonance to reduce the s-wave scattering length to smaller values such that $\xi_0 \sim D$ or make the packing of the vortices in the lattice so dense that the $\xi_0 \sim D$ condition is satisfied. One can reasonably expect here that the thin vortices to show up in the absence of thick vortices and some already trapped quantized angular momenta.

\par
\section{Discussion}

Note that in arriving at the new class of vortex solution we have used here the already existing method of fixing the vortex core structure, in the presence of non-local corrections. The variational procedure then is employed to make this vortex core smoothly evolve to the uniform density far field. As opposed to Pad\'e type approximate solution being used as a variational ansatz, which is standard, our variational analysis(practically for all vortices) is using a well determined solution of the core of the vortices at small $r$ up to next to leading order approximation. As we have already shown in the section named 'Standard Theory', at the leading order, at small $r$, near the core of a vortex, the conventional theory based on the dynamics cannot give a length scale selection and works fine for the thick vortices. However, in order to go to the sub-leading corrections, one must take the non-local correction into account which we have done and have determined the solution for the core of the vortex from the dynamics itself where the appropriate length scale shows up. Then, while making this core of the vortex recover to the uniform field, far away from the core, we have used variation to have a smooth transition. Thus, our variational approach is just not based entirely on guessed ansatz, rather, the most important part of the solution is obtained from the dynamics and then it is extrapolated to the far field smoothly using variation.

\par
Note that the results involving higher order vortices resulting from the higher order terms of the expansion although are coming out through a general procedure, but, may not all be physically accessible because of involvement of higher derivative i.e. smaller length scales. However, the first correction term should always be there since a Laplacian is already present in the local model. So, the result involving the lowest order vortex which is most stable is there.

\par
A consistent mathematical analysis following well established methods of theoretically capturing vortex solutions, indicate here the presence of thin vortex solutions due to non-local interactions. The non-local interaction correction taken in the present context and in other previous works \cite{nonlocal_pethick} presents small length scales on top of that captured by the mean field GP dynamics. The origin of such fluctuations probably lies in the quantum effects not captured by the local GP dynamics. In that sense, the dynamics of these thin vortices would possibly manifest quantum fluctuations in the system which are required for vortex lattice melting and also important in the context of superfluid turbulence that has attracted a lot of attention lately \cite{superfluid_turbulence_kozik}. Apart from the possible experimental situation where the thick vortex breaks down, which we have identified as a plausible situation where our thin vortex solution can be realized, a quench through the critical point and subsequent formation of vortex-antivortex pairs through Kibble-Zurek \cite{kibble_zurek_damski} mechanism could also be a possible procedure to realize such a vortex. If experimentally realized, these objects can possibly demonstrate quantum fluctuations in vortices which involves rich physics.

\vspace{6mm}
\begin{acknowledgements}
Abhijit Pendse would like to acknowledge the support provided by the Council of Scientific and Industrial Research(CSIR), India. AP and AB would like to acknowledge very useful suggestions by anonymous referees which has helped us put this work in proper context.
\end{acknowledgements}

\par
AP and AB have made equal contributions to this research.

\bibliography{references}

\end{document}